\documentclass[acmsmall, nonacm]{acmart}

\usepackage{algpseudocode}
\usepackage{graphicx}
\usepackage{textcomp,comment}
\usepackage{xcolor}
\usepackage{mathtools}

\usepackage{bbm}
\usepackage{mathrsfs}
\usepackage{float}
\usepackage{setspace}
\usepackage{epstopdf}

\usepackage{textcomp}
\usepackage[linesnumbered,ruled,vlined]{algorithm2e}
\usepackage{dsfont}
\newtheorem{claim}{Claim}
\newtheorem{assumption}{Assumption}

\usepackage{tikz}

\renewcommand{\hat}{\widehat}
\renewcommand{\bar}{\overline}

\AtBeginDocument{%
  \providecommand\BibTeX{{%
    \normalfont B\kern-0.5em{\scshape i\kern-0.25em b}\kern-0.8em\TeX}}}

\setcopyright{acmcopyright}
\copyrightyear{2025}
\acmYear{2025}
\acmDOI{XXXXXXX.XXXXXXX}

\acmJournal{JACM}
\acmVolume{0}
\acmNumber{0}
\acmArticle{0}
\acmMonth{5}

\begin{document}

\title{Mean Field Analysis of Blockchain Systems}

\author{Yanni Georghiades}
\email{yanni.georghiades@utexas.edu}
\orcid{0000-0002-3244-0815}
\affiliation{%
  \institution{The University of Texas at Austin}
  \city{Austin}
  \state{Texas}
  \country{USA}
}

\author{Takashi Tanaka}
\email{tanaka16@purdue.edu}
\affiliation{%
  \institution{Purdue University}
  \city{West Lafayette}
  \state{Indiana}
  \country{USA}
}

\author{Sriram Vishwanath}
\email{sriram@utexas.edu}
\orcid{0000-0003-3112-4885}
\affiliation{%
  \institution{The University of Texas at Austin}
  \city{Austin}
  \state{Texas}
  \country{USA}
}

\renewcommand{\shortauthors}{Georghiades, et al.}

\begin{abstract}
We present a novel framework for analyzing blockchain consensus mechanisms by modeling blockchain growth as a Partially Observable Stochastic Game (POSG) which we reduce to a set of Partially Observable Markov Decision Processes (POMDPs) through the use of the mean field approximation. This approach formalizes the decision-making process of miners in Proof-of-Work (PoW) systems and enables a principled examination of block selection strategies as well as steady state analysis of the induced Markov chain. By leveraging a mean field game formulation, we efficiently characterize the information asymmetries that arise in asynchronous blockchain networks.

Our first main result is an exact characterization of the tradeoff between network delay and PoW efficiency--the fraction of blocks which end up in the longest chain. We demonstrate that the tradeoff observed in our model at steady state aligns closely with theoretical findings, validating our use of the mean field approximation.

Our second main result is a rigorous equilibrium analysis of the Longest Chain Rule (LCR). We show that the LCR is  a mean field equilibrium and that it is uniquely optimal in maximizing PoW efficiency under certain mild assumptions. This result provides the first formal justification for continued use of the LCR in decentralized consensus protocols, offering both theoretical validation and practical insights.

Beyond these core results, our framework supports flexible experimentation with alternative block selection strategies, system dynamics, and reward structures. It offers a systematic and scalable substitute for expensive test-net deployments or ad hoc analysis. While our primary focus is on Nakamoto-style blockchains, the model is general enough to accommodate other architectures through modifications to the underlying MDP.

Together, these contributions lay a rigorous foundation for studying and designing blockchain consensus mechanisms. Our work provides a structured, adaptable, and theoretically grounded lens through which to analyze decentralized systems at scale.

\end{abstract}



\maketitle

\section{Introduction}
\label{sec:intro}
Blockchains have fundamentally transformed the design of decentralized systems by providing consensus mechanisms that do not rely on trusted intermediaries. Among these, the Longest Chain Rule (LCR), popularized by Bitcoin, has become the de facto block selection strategy in Proof-of-Work (PoW) systems. Despite its empirical success and widespread adoption, a rigorous theoretical justification for LCR's optimality has remained elusive. Simultaneously, understanding how network delays and block generation rates impact consensus stability in PoW networks remains a critical challenge—particularly given the increasing need for scalable and efficient blockchain architectures.

In this work, we provide a formal computational framework that addresses these fundamental gaps. We model blockchain growth as a Partially Observable Stochastic Game (POSG) which we reduce to a set of Partially Observable Markov Decision Processes (POMDPs) through the use of the mean field approximation.  
We propose a tractable solution methodology based on iteratively solving successive POMDPs under the fully observable value approximation heuristic. Our approach captures the dynamics of block generation, propagation delays, and information asymmetries between agents operating in decentralized networks, and it allows for steady state analysis of the induced Markov chain under any set of system parameters or equilibrium policy.

Our contributions are as follows:

\begin{description}
     \item[POMDP-Based Blockchain Growth Model:] We present a novel framework that models blockchain growth as a set of POMDPs, providing a structured and formalized approach to analyzing block generation and selection dynamics when agents only receive partial observations of the underlying system state.

     \item[Flexible Framework for Block Selection Strategies:] Our model facilitates the systematic testing and comparison of various block selection strategies, overcoming the limitations of manual analysis and expensive test-net implementations. It also supports modifications to block propagation and generation dynamics, reward structures, and blockchain architectures.
     \item[Characterization of Delay vs. PoW-Efficiency Tradeoff:] Through steady state analysis, we establish an exact characterization of the tradeoff between network delay and PoW efficiency--the fraction of blocks which end up as part of the longest chain--derived from our model’s equilibrium behavior. Our analysis shows that this tradeoff quantitatively aligns with established analytical models of blockchain throughput \cite{georghiades2022scalable, pass2017analysis}. This result affirms usage of the mean field approximation in predicting emergent behaviors in large-scale PoW systems and validates our framework as a robust tool for studying blockchain consensus.

    \item[Optimality of the LCR:] We prove that the LCR is a mean field equilibrium strategy and that it is uniquely optimal in maximizing PoW efficiency under certain mild assumptions as outlined in Section \ref{sec:model}. To our knowledge, this is the {\bf first formal proof of the LCR’s optimality in decentralized consensus} under a well-defined policy class, offering a theoretical foundation for its continued use in distributed consensus systems.

    \item[Adaptability Beyond Nakamoto-Style Blockchains:] While our primary focus in this work is on Nakamoto-style blockchains, the framework is adaptable to other architectures through modifications to the underlying state space and transition functions, making it a versatile tool for broader blockchain research.
\end{description}

Together, these contributions provide a novel lens for understanding blockchain performance and strategy. Our framework not only bridges the gap between empirical heuristics and theoretical analysis, but also offers a scalable, generalizable foundation for exploring a broader class of protocols beyond Nakamoto-style blockchains.

\section{Related Work}
\label{sec:related}
This section reviews key contributions in three relevant areas: Bitcoin and other blockchain implementations, single- and multi-agent dynamic games, and mean field games. These domains intersect in the analysis and modeling of blockchain systems, and each plays a critical role in the development of the simulation framework we propose.

\subsection{Bitcoin and Other Blockchain Implementations}
Bitcoin, introduced by Satoshi Nakamoto in 2008, was the first implementation of a decentralized cryptocurrency that leverages blockchain technology for secure, peer-to-peer transactions without relying on centralized intermediaries \cite{nakamoto2008bitcoin}. The consensus mechanism used in Bitcoin is called Proof of Work (PoW), where miners compete to solve computationally intensive puzzles. Once a valid solution is found, the miner appends a block of transactions to the blockchain, and the chain is extended. The LCR, a heuristic used in Bitcoin, ensures that nodes in the network attempt to append to the chain with the most cumulative work. However, while the LCR has demonstrated empirical success, its theoretical foundations are not yet fully understood. There is no formal proof that it is optimal or even that it is the best available chain selection strategy under all conditions.

Beyond Bitcoin, numerous other blockchain systems have emerged, such as Ethereum \cite{buterin2013ethereum}, Litecoin \cite{reed2017litecoin}, and Zcash \cite{hopwood2016zcash}, each featuring different consensus mechanisms and design trade-offs. Ethereum, for instance, has been transitioning from a PoW system to a Proof of Stake (PoS) mechanism, where miners are replaced by validators who are selected to propose and validate new blocks based on the amount of cryptocurrency they hold and are willing to ``stake.'' PoS eliminates the energy-intensive mining process of PoW and introduces new challenges in chain selection and consensus strategies. Other blockchain systems explore alternative architectures like Directed Acyclic Graphs (DAGs) used by IOTA, which move away from traditional linear blockchain structures.

Despite the diversity of blockchain implementations, a common challenge remains: evaluating the security, efficiency, and optimality of different blockchain architectures and block selection strategies. 
Existing studies typically take one of two approaches: i) they rely on empirical measurement of real world systems or simulators, allowing for observation (but not formal analysis) of a wide variety of blockchain properties \cite{decker2013information, miller2015shadow}; or ii) they involve ad-hoc theoretical models which are specific to individual blockchain architectures and a limited set of blockchain properties \cite{pass2017analysis, georghiades2023majority, georghiades2022scalable, garay2024bitcoin, dembo2020everything}. 
We propose a rigorous, generalizable framework based on Partially Observable Markov Decision Processes for studying the stationary behavior of blockchain systems.  
Unlike simulation-based methods, our approach captures complex information asymmetries and edge cases that arise in asynchronous networks, and it can be used to determine an equilibrium policy for any set of system parameters. 
Our framework can be extended to support formal analysis of many blockchain properties and architectures.

\subsection{Markov Decision Processes (MDPs)}

Markov Decision Processes (MDPs) are a well-established framework in decision theory and reinforcement learning for modeling decision-making in stochastic environments \cite{puterman2014markov}. 
An MDP is a tuple $(\mathcal{S}, \mathcal{A}, T, R)$, where $\mathcal{S}$ is a finite state space; $\mathcal{A}$ is a finite set of actions; $T(S_t, a, S_{t+1})$ is the probability of transitioning from state $S_t$ to state $S_{t+1}$ under action $a$; and $R(S_t, a, S_{t+1})$ is the reward granted when transitioning from state $S_t$ to state $S_{t+1}$ under action $a$.

MDPs provide a structured approach to analyzing sequential decision-making strategies and are particularly useful for problems where the outcome of an action is uncertain and future states depend only on the current state and the chosen action.
The process of adding new blocks to a blockchain can be seen as a sequential decision-making problem in which miners must choose the best block to append to. Each selection influences the future state of the chain and affects the rewards received, making MDPs a natural fit for analyzing the incentives and strategies that drive blockchain growth.

Existing research has used MDPs to study mining behaviors and security vulnerabilities, such as selfish mining and block withholding attacks \cite{jofre2021convergence, nayak2016stubborn}. However, these studies largely focus on specific attack vectors or narrow aspects of blockchain design. Our work extends this approach by providing a general framework for blockchain growth, enabling the evaluation and steady state analysis of not only attack vectors but also chain selection rules, block propagation dynamics, and reward structures.

\subsection{Stochastic games}
A stochastic game generalizes a Markov Decision Process (MDP) to a multi-agent setting, where transitions and rewards depend on the collective actions of all players \cite{shapley1953stochastic}. Unlike in an MDP, where a single agent optimizes its policy against a fixed environment, a stochastic game models strategic interactions among multiple decision-makers, each seeking to maximize their individual rewards.
Concretely, a stochastic game $(\mathcal{S}, \mathcal{N}, \mathcal{A}, T, R)$ differs from an MDP in that it considers a set of $N$ agents and the actions and rewards are considered jointly over all agents. 

In the context of blockchains, stochastic games have been used to show that the LCR is a Nash equilibrium of the mining game \cite{kiayias2016blockchain}.
However, this model is limited in that it considers only two miners and does not consider the impact of network propagation delays.
Our model considers an arbitrarily large number of agents and characterizes the relationship between propagation delays and the partial observations of each agent.

\subsection{Partially Observable Stochastic Games and Markov Decision Processes}

Partially Observable Stochastic Games (POSGs) \cite{albrecht2024multi} and Partially Observable Markov Decision Processes (POMDPs) \cite{krishnamurthy2016partially} extend stochastic games and MDPs by accounting for uncertainty in state observations. In a standard stochastic game or MDP, agents have full knowledge of the system's state. However, in many real-world scenarios, including blockchain systems, the true state of the environment is not fully observable. 
In addition to the standard elements of stochastic games and MDPs defined above, POSGs and POMDPs define a finite set of observations $\Omega^i$ for each agent and a function $\mathcal{O}^i(S_t, a, \Omega^i_t)$ which defines the probability of each observation. 
That is, agents receive only partial or noisy observations of the system state, making decision-making significantly more complex.
To navigate uncertainty, each agent maintains a belief state—a probability distribution over possible states—which is updated dynamically as new observations arrive.

POMDPs have been widely used in robotics, finance, healthcare, and artificial intelligence, where decision-making under uncertainty is critical. Their application to blockchain systems, however, remains underexplored. In the blockchain context, miners often operate under partial information due to network delays or adversarial behavior. For example, a miner may not have immediate knowledge of the full transaction pool or the current longest chain, requiring them to make block selection decisions based on incomplete information.

\subsection{Mean Field Games}

Mean Field Games (MFGs) \cite{Caines2018, lasry2007mean} are used to model settings with a large number of interacting agents, offering a scalable framework for analyzing strategic decision-making in multi-agent systems. Multi-agent games often become computationally intractable as the number of players increases due to the exponential growth in joint action spaces and state transitions. MFGs circumvent this issue by introducing a mean field approximation, where each agent interacts not with individual opponents but with the aggregate effect of the entire population.

Recently, MFGs have been used to model the wealth distribution of cryptocurrency miners in order to study wealth heterogeneity over time \cite{li2024mean} . 
In our framework, we use MFGs to model information asymmetries in the block graph that arise from an asynchronous communication network.

\section{System Model}
\label{sec:model}
We consider a discrete-time blockchain system with $N$ rational agents, where each agent independently attempts to generate and append blocks to the blockchain. To model information asymmetries inherent to real-world blockchain networks, agents make decisions based solely on their local block graph—a subgraph of the global block graph containing only the blocks they have received up to that point in time.

\subsection{Time Units in the Model}

Our system operates with two distinct units of time, which allow us to construct a flexible yet computationally tractable framework for modeling block generation and block propagation.

\subsubsection{Time Steps} 
The smallest unit of time in our model, with a fixed duration. Time steps drive both block generation and block propagation, which we model as Bernoulli processes. In each time step: i) a block is generated with some probability, and ii) each agent independently receives any block from the global block graph with some probability.

However, depending on system parameters, the number of time steps required for a block to be generated can be large\footnote{For instance, a time step might correspond to a single execution of a Proof-of-Work function.}. Given this potential variance, modeling state transitions at the time step level is computationally infeasible.
As such, we introduce a second measure of time. 

\subsubsection{Block Steps}
A block step represents a number of time steps required for a block to be generated. Since block generation is stochastic, the number of time steps in a block step is random. State transitions occur at the granularity of block steps, meaning each agent selects exactly one action per block step.
Block steps are generally referred to as $t$ or time, as a block step is the unit of time for our game.

\subsection{Block Generation}
Block generation follows a Bernoulli process. 
In each time step, a block is generated with probability $\alpha$, a system parameter that can be adjusted to reflect different block propagation conditions.
Each agent has an equal probability of generating a block in any given time step, and upon generating a block, the agent immediately broadcasts it to all other agents.
At most one block is generated per time step\footnote{This can be assumed without loss of generality, as time steps can be made arbitrarily small, ensuring that the probability of multiple blocks being generated in a single time step is negligible.}.
By structuring block generation and propagation in this way, our model provides a simple yet robust foundation for analyzing blockchain dynamics under various conditions.

Block generation is characterized by the following equations. 
If $k_{\alpha}$ is a random variable denoting the number of time steps in a block step, then the probability that there are $k$ time steps in a block step is

\begin{equation}
    P(k_{\alpha} = k) = \begin{cases}
    (1 - \alpha)^{k-1} \alpha &  \text{if  }  k \geq 1 \\
    0 & \text{else}
    \end{cases}.
\end{equation}

More generally, if $k_{\alpha}^y$ is the number of time steps in $y$ block steps, then the probability that there are $k$ time steps in $y$ block steps is
\begin{equation}
    P(k_{\alpha}^y = k) = \begin{cases}
    {k - 1 \choose k- y} (1 - \alpha)^{k-y}  \alpha^y &  \text{if  }  k \geq y \\
    0 & \text{else}
    \end{cases}.
\end{equation}

\subsection{Block Propagation Model}

We model block propagation as a collection of \textit{independent and identically distributed (i.i.d.)} stochastic processes, with each process corresponding to a unique \textit{(block, agent)} pair. At each time step, for any block \( x \) that an agent has not yet received, the agent receives \( x \) with probability \( \delta \).
This model assumes a fully connected communication graph, meaning any agent can receive a block from any other agent.

The probability that an agent receives a block depends on the number of block steps that have elapsed since the block was generated. Let \( H_{\alpha, \delta} \) be the number of block steps required for an agent to receive a block. Then, the probability that an agent receives a block within \( h  \) block steps is given by

\begin{equation}
    P(H_{\alpha, \delta} \leq h) = \sum_{k=1}^{\infty} (1 - (1 -\delta)^k) P(k_{\alpha}^{h} = k).
\end{equation}

The probability that an agent receives the block at exactly block step \( h \) is then

\begin{equation}
    P(H_{\alpha, \delta} = h) = P(H_{\alpha, \delta} \leq h) - P(H_{\alpha, \delta} \leq h-1).
\end{equation}

Since the process governing block propagation is \textit{memoryless}, the probability \( P(H_{\alpha, \delta} = h) \) remains the same for any block step at which the agent has not yet received the block. 

\section{Mean Field Game  for Equilibrium Dynamics in Blockchain Growth}
\label{sec:mdp}

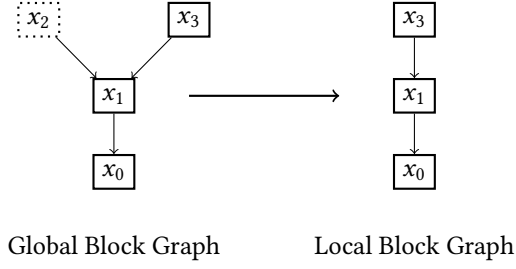
\begin{figure}
\begin{center}
\begin{tikzpicture}
    \tikzstyle{block}=[thick,draw=black,rectangle]
    \tikzstyle{block2}=[thick,draw=black,rectangle,dotted]

    \node[block] at (2,1) 
    (4_3_1_1__0) {$x_0$};
    \node[block] at (2,2) 
    (4_3_1_1__1) {$x_1$};
    \node[block2] at (1,3) 
    (4_3_1_1__2) {$x_2$};
    \node[block] at (3,3) 
    (4_3_1_1__3) {$x_3$};
    \node at (2,0) {Global Block Graph};
    \draw[<-] (4_3_1_1__1) edge (4_3_1_1__3)
              (4_3_1_1__1) edge (4_3_1_1__2)
              (4_3_1_1__0) edge (4_3_1_1__1);

    \draw[->, thick](3,2) to (5,2);

    \node[block] at (6,1) 
    (3_2__0) {$x_0$};
    \node[block] at (6,2) 
    (3_2__1) {$x_1$};
    \node[block] at (6,3) 
    (3_2__2) {$x_3$};
    \node at (6,0) {Local Block Graph};
    \draw[<-] (3_2__0) edge (3_2__1)
              (3_2__1) edge (3_2__2);
    
\end{tikzpicture}
\end{center}
\caption{An example of how a local block graph can be derived from a local block graph. In this example, the agent has received blocks $x_0, x_1$, and $x_3$, so their local block graph is the connected subgraph containing these blocks.}
\label{fig:global_to_local}
\end{figure}

In general, blockchain growth can be formulated as the POSG
\[
\mathcal{M}_{\texttt{POSG}} = (\mathcal{S}, \mathcal{N}, \mathcal{A}, T, R, \Omega, \mathcal{O}), 
\]
where a state $S_t \in \mathcal{S}$ is represented by the tuple 
\[
S_t = \big(G_t, \{B^i_t\}_{i = 1,2,\dots,N}, \{O^i_t\}_{i = 1,2,\dots,N}\big).
\]
Here, $G_t$ is the global block graph, $B^i_t$ is a \(\{0,1\}\)-vector indicating which blocks from \( G_t \) have been received by agent \( i \), and $O^i_t$ is a \(\{0,1\}\)-vector indicating which blocks from \( G_t \) were generated by agent $i$.
Rather than observing $S_t$, an agent $i$ observes only $\Omega^i_t$, which corresponds to the set of blocks they have received as of time $t$. 
This observation is referred to as their \textit{local block graph}, as it corresponds to the block graph they observe in their local view.
An example of a local block graph is shown in Figure \ref{fig:global_to_local}.

In every block step, the action taken by each agent corresponds to selecting a block from their local block graph that they want to append to.
As such, we refer to a policy mapping a local block graph to an action as a \textit{local policy} $\bar{\pi}$. 
Later, we will make use of a \textit{full policy} $\pi$, which maps a full state to an action as in a fully observable stochastic game.

This model conforms closely to blockchain growth in practice, as it fully captures the system state at any given time. 
However, since the cardinality of the state space of $\mathcal{M}_{\texttt{POSG}}$ grows exponentially with the number of agents, it is difficult to analyze such a game directly. 
In order to analyze blockchain growth in a computationally tractable manner, we adopt the mean field approximation.

\subsection{The Mean Field Approximation}
Modeling blockchain growth as a mean field game requires the following assumptions. 

\begin{assumption}
\label{ass:large}
    The number of agents is large.
\end{assumption}

\begin{assumption}
\label{ass:individual}
    Each agent's individual impact on the growth of the blockchain is negligible.
\end{assumption}

\begin{assumption}
\label{ass:equal}
    Agent identities are indistinguishable, and all agents have an equal probability of generating a block in a given block step.
\end{assumption}

Assumption \ref{ass:large} is quite reasonable in practice, as it is generally agreed that a larger number of miners leads to a greater degree of decentralization in cryptocurrencies. 
In practice, Assumptions \ref{ass:individual} and \ref{ass:equal} are not universal operating conditions, as some cryptocurrencies have individual miners or mining pools which control a significant fraction of the computational power, but we discuss in Section \ref{sec:conclusions} how these assumptions might be relaxed to enable the study of such systems.
However, for simplicity, we consider only a homogeneous distribution of computational power in this work. 

Under the mean field approximation, blockchain growth can be reduced to the single-agent optimal control problem
\[
\mathcal{M}_{\texttt{POMDP}} = (\mathcal{S}, \mathcal{A}, T, R, \Omega, \mathcal{O}), 
\]
which is similar to $\mathcal{M}_{\texttt{POSG}}$ except that the states, actions, rewards, and observations correspond only to a single representative agent (which we will refer to indiscriminately as the RA, the representative agent, or agent \(j\), when an index is required) instead of a set of $N$ agents. 
Under this model, the observations of the remaining Non-Representative Agents (NRAs) are characterized in distribution by the mean field variable $\mu$, allowing them to be represented efficiently without any impact on the size of the state space.
The NRAs are assumed to follow a fixed policy $\bar{\pi}$, and the solution to the resulting POMDP can be interpreted as the best response of the RA when the NRAs follow local policy $\bar{\pi}$ and have local block graphs distributed according to $\mu$.

At this point, we move away from standard notation in order to depict our model and solution method more clearly later on. 
First, note that because $S_t$ contains the global block graph and a representation of which blocks the RA has received, $\Omega_t$ can be computed directly as a function of $S_t$. 
Second, because solving POMDPs is in itself computationally challenging, we adopt the well-known \textit{fully observable value approximation} heuristic (terminology adopted from \cite{kochenderfer2022algorithms}). 
In this approach, the value function of the POMDP, which is difficult to compute, is approximated using the value function of the (fully observable) MDP, which can be found efficiently \cite{hauskrecht2000value}.
We describe this heuristic in more detail in Section \ref{sec:mfa}.
With this in mind, the variables $\Omega$ and $\mathcal{O}$ are redundant and can safely be dropped from our notation. 
Additionally, because we are solving only for the best response of the RA, the NRAs can be treated as a part of the system dynamics, where each pair $(\bar{\pi}, \mu)$ corresponds to a different set of dynamics which impact the transition and reward functions. 
As such, we adopt the custom notation 
\[
\mathcal{M}_{\bar{\pi}} = (\mathcal{S}, \mathcal{A}, T_{\bar{\pi}}, R_{\bar{\pi}}, \mu) 
\]
to refer to the POMDP corresponding to NRA policy $\bar{\pi}$ and mean field distribution $\mu$.
We represent our multi-agent game as $\mathcal{M}$, which corresponds to the set of all single-agent POMDPs $\mathcal{M}_{\bar{\pi}}$.
In order to determine equilibrium policies of $\mathcal{M}$, we iteratively solve successive POMDPs until a fixed point is reached. 

In the remainder of this section, we provide greater detail on each component of $\mathcal{M}_{\bar{\pi}}$.

\subsection{States}

The state at time $t$ takes the form 
\[
S_t = (G_t, B_t, O_t),
\]
where $G_t$ is the global block graph, $B_t$ is a \(\{0,1\}\)-vector indicating which blocks from \( G_t \) have been received by the RA, and $O_t$ is a a \(\{0,1\}\)-vector indicating which blocks from \( G_t \) were generated by the RA (which is necessary for computing the reward function).
For simplicity, we overload vector indexing notation such that if a block $x$ has index $k$, then $B_t[x]$ and $B_t[k]$ are used interchangeably to refer to the value of $B_t$ at index $k$.
For a block $x \in G_t$, $O_t[x] = 1$ if and only if the representative agent generated $x$; similarly, $B_t[x] = 1$ if and only if the agent has received $x$ as of time $t$.

As $\mathcal{M}$ is partially observable, agents are not assumed to have full knowledge of the state at any given time. 
Instead, agents must make decisions based off of their local block graph, which contains only the blocks from the global block graph the agent has received at the current state. 
The \textbf{local block graph} of an agent \( i \), denoted \( L^i_t \), is defined as follows: if agent \( i \) has not received the root vertex of $G_t$, then \( L^i_t \) is the null graph; otherwise, \( L^i_t \) is the largest connected subgraph of $G_t$ that contains its root block.
Figure \ref{fig:global_to_local} shows an example of how a local block graph can be converted to a local block graph.

$L^j_t$ is the local block graph of the RA at time $t$ and can be computed directly from the state, as $B_t$ specifies precisely which blocks from $G_t$ the RA has received. 
However, the local block graph of an agent $i \neq j$ is known only in distribution, where 
\[
P(L^i_t = l) = \mu(l, G_t).
\]

\subsection{Actions}
At time $t$, the RA selects an action $a_t \in \mathcal{A}$, where the action space is the set of blocks in their local block graph. 
When any agent generates a block, it is appended to the block $G_t$ corresponding to their action. 
An agent's available actions are restricted to the blocks in their local block graph; if their local block graph is the null graph, then no action is available\footnote{This modeling decision mimics the scenario in which a miner or validator is so far out of sync with the network that their decision has negligible impact on the evolution of the blockchain.}.

Agents are assumed to have incomplete knowledge of the state, so we are primarily interested in the \textit{local policy}  of each agent 
\[
\bar{\pi} : \mathcal{G} \to \mathcal{A},
\]
where $\bar{\pi}_i(L^i)$ is the action taken by agent $i$ and is dependent only on knowledge of their local block graph.
However, in Section \ref{sec:compute_mfe} we also make use of the \textit{full policy}, which is given by the decision rule 
\[
\pi: \mathcal{S} \to \mathcal{A},
\]
where $\pi_i(S_t)$ is the action agent $i$ would take under complete knowledge of $S_t$. 

\subsection{Transition Probabilities}
A state transition occurs exactly once per block step, meaning one block is appended to the global block graph per state transition.
Under a given local policy $\bar{\pi}$, the probability of transitioning from state $S_t$ to $S_{t+1}$ when the representative agent plays action $a_t$ can be decomposed into the component probabilities

\[
T_{\bar{\pi}}(S_t,a_t, \mu, S_{t+1}) =  P(G_{t+1} \mid S_t, a_t, \mu, \bar{\pi}) \, P(O_{t+1} \mid S_t, a_t, \mu, \bar{\pi}, G_{t+1}) \, P(B_{t+1} \mid S_t, a_t, \mu, \bar{\pi}, G_{t+1}, O_{t+1}),
\]

where $P(G_{t+1} \mid S_t, a_t, \mu, \bar{\pi})$, $P(O_{t+1} \mid S_t, a_t, \mu, \bar{\pi}, G_{t+1})$, and $P(B_{t+1} \mid S_t, a_t, \mu, \bar{\pi}, G_{t+1}, O_{t+1})$ are the probabilities of transitioning to global block graph $G_{t+1}$, ownership vector $O_{t+1}$, and received block vector $B_{t+1}$, respectively.
In Section \ref{sec:mfa}, we describe a methodology for computing each component probability individually for Nakamoto-style blockchains.

\subsection{Rewards}
Although there are a variety of possible reward models, we define a generic reward function which rewards agents for generating blocks. 
In Nakamoto-style cryptocurrencies, it is common for rewards to be issued for generating blocks along some critical path (in Bitcoin the critical path is the chain of blocks in the global block graph with the most cumulative work). 
Rewards are generally distributed to miners after some condition is met, such as a sufficiently large number of subsequent block confirmations. 
In deference to these general constraints, we define the reward for the representative agent

\[
R_{\bar{\pi}}(S_t, a, S_{t+1}, \mu) = r Q(S_t,a,S_{t+1}, \mu, \bar{\pi}), 
\]

where $Q(S_t,a,S_{t+1}, \mu, \bar{\pi})$ is the number of critical path blocks for which the agent should receive rewards, and $r$ is the associated reward for a block.
There are a variety of ways to define $Q$, and in Section \ref{sec:mfa} we define $Q$ in the context of Nakamoto-style blockchains.  

\subsection{Equilibrium Concept}
\label{sec:eq_c}
The solution concept we desire is a mean field equilibrium \cite{adlakha2015equilibria}. 
That is, a policy $\bar{\pi}^*$ which is a Nash equilibrium of $\mathcal{M}$ when each agent optimizes their actions against the long-run average behavior of the other agents. 
Concretely, if $\mu_{\bar{\pi}^*}$ is the mean field distribution of $\mathcal{M}_{\bar{\pi}^*}$ at steady-state, then the mean field equilibrium policy is characterized by the Bellman equation
\[
\bar{\pi}_j^*(L^j_t) = \arg \max_a \sum_{s} T_{\bar{\pi}^*}(S_t,a,s,\mu_{\bar{\pi}^*})[R_{\bar{\pi}^*}(S_t,a,s,\mu_{\bar{\pi}^*}) + \gamma V^*(s)],
\]
where $\gamma \in [0,1]$ is a discount factor and
\[
V^*(s) = \max_a \sum_{s'} T_{\bar{\pi}^*}(s,a,s',\mu_{\bar{\pi}^*})[R_{\bar{\pi}^*}(s,a,s',\mu_{\bar{\pi}^*}) + \gamma V^*(s')]
\]
is the expected discounted future rewards for state $s$.
Under this notation, an equilibrium is characterize
Our method of solving for $\bar{\pi}^*$ follows an iterative best-response learning process \cite{lasry2007mean, Caines2018}:

\begin{enumerate}
\item We first determine the optimal policy for the RA under the assumption that the NRAs follow a predefined policy.
\item Next, we update the policy of the NRAs to match the optimal policy learned by the RA.
\item We repeat this process, iteratively computing the RA’s best response to the updated NRA policy until the strategies converge.
\end{enumerate}

Convergence indicates that an equilibrium strategy has been reached, meaning no agent has an incentive to unilaterally deviate. This iterative approach allows us to approximate optimal decision-making in large-scale blockchain environments where direct analytical solutions are impractical.
In Section \ref{sec:compute_mfe}, we describe in detail how $\mu_{\bar{\pi}^*}$ can be determined.

\section{Equilibrium Policies of a Nakamoto-Style Blockchain}
\label{sec:mfa}
In this section, we define a state space, transition probability function, and reward function which correspond to a Nakamoto-style blockchain. 
We then define a custom algorithm which solves for a mean field equilibrium of the resulting game under any system parameters.
The cardinality of the state space is large due to the combinatorial nature of the block graph and its associated vectors, so we implement a number of  techniques to limit the size of the state space. 
We provide equations for determining transition probabilities, including an iterative algorithm for computing the mean field distribution, and we define a concrete reward function. 

\subsection{Global Block Graph}
The first component of the state is the global block graph $G_t$, a directed graph where each vertex is a block and each edge is a reference to a prior block. 
We assume that blocks are functionally identical, meaning all blocks are equally difficult to generate and are worth the same reward, but that blocks can still be distinguished from one another (e.g., by their hash value). 
Although no single agent's local block graph necessarily contains every block in $G_t$, it is not possible for an agent to have a block in their local block graph which is not in $G_t$. 
Recall that the local block graph $L^j_t$ of the representative agent can be computed from $S_t$, while the local block graphs of the remaining agents are characterized in distribution by $\mu$ and cannot be computed directly. 

\subsubsection{Graph Isomorphism}
In the Nakamoto-style blockchains we consider, the root vertex of any graph $G \in \mathcal{G}$ has an outdegree of 0, and every other vertex has an outdegree of 1. 
That is, each block other than the root has a directed edge to exactly one prior block, although multiple blocks may reference the same prior block. 
We refer to a directed graph with this structure as a \emph{Nakamoto graph}\footnote{Our model could be extended to analyze alternative blockchain structures, such as DAG-based blockchains, by defining a different set of isomorphism classes for the block graph.}.
There are many groupings of Nakamoto graphs which share the same topology, and under the assumption that all blocks are functionally identical, any two graphs with the same topology will provoke the same agent response (outside of considerations of block ownership).
Thus, we can restrict $\mathcal{G}$ to contain only the isomorphism classes of a Nakamoto graph, i.e., we need consider only one graph per topology rather than all possible combinations of blocks and edges. 

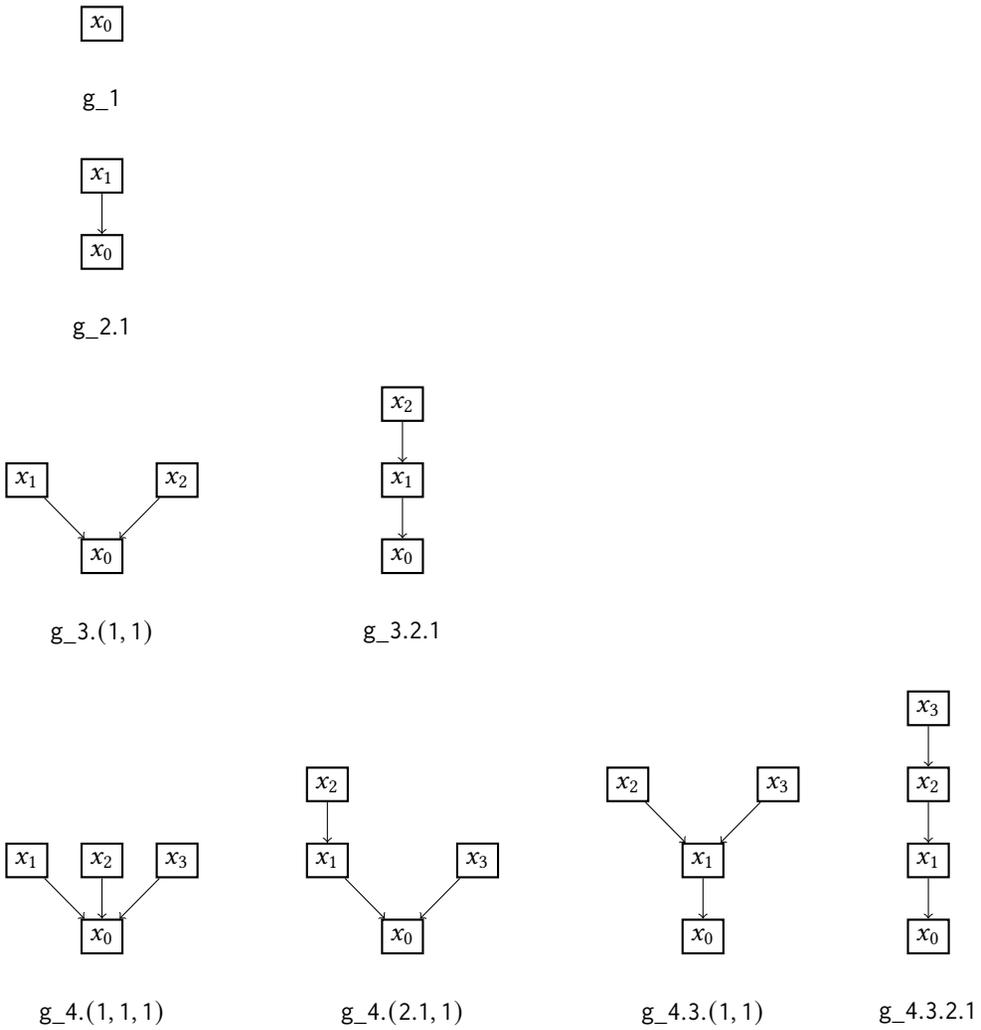
\begin{figure}

\begin{center}
\begin{tikzpicture}
    \tikzstyle{block}=[thick,draw=black,rectangle]

    \node[block] at (1,1) 
    (4_1_1_1__0) {$x_0$};
    \node[block] at (0,2) 
    (4_1_1_1__1) {$x_1$};
    \node[block] at (1,2) 
    (4_1_1_1__2) {$x_2$};
    \node[block] at (2,2) 
    (4_1_1_1__3) {$x_3$};
    \node at (1,0) {$\mathtt{g\_4.(1,1,1)}$};
    \draw[<-] (4_1_1_1__0) edge (4_1_1_1__1)
              (4_1_1_1__0) edge (4_1_1_1__2)
              (4_1_1_1__0) edge (4_1_1_1__3);

    \node[block] at (5,1) 
    (4_2_1__0) {$x_0$};
    \node[block] at (4,2) 
    (4_2_1__1) {$x_1$};
    \node[block] at (6,2) 
    (4_2_1__2) {$x_3$};
    \node[block] at (4,3) 
    (4_2_1__3) {$x_2$};
    \node at (5,0) {$\mathtt{g\_4.(2.1,1)}$};
    \draw[<-] (4_2_1__0) edge (4_2_1__2)
              (4_2_1__1) edge (4_2_1__3)
              (4_2_1__0) edge (4_2_1__1);

    \node[block] at (9,1) 
    (4_3_1_1__0) {$x_0$};
    \node[block] at (9,2) 
    (4_3_1_1__1) {$x_1$};
    \node[block] at (8,3) 
    (4_3_1_1__2) {$x_2$};
    \node[block] at (10,3) 
    (4_3_1_1__3) {$x_3$};
    \node at (9,0) {$\mathtt{g\_4.3.(1,1)}$};
    \draw[<-] (4_3_1_1__1) edge (4_3_1_1__3)
              (4_3_1_1__1) edge (4_3_1_1__2)
              (4_3_1_1__0) edge (4_3_1_1__1);

    \node[block] at (12,1) 
    (4_3_2__0) {$x_0$};
    \node[block] at (12,2) 
    (4_3_2__1) {$x_1$};
    \node[block] at (12,3) 
    (4_3_2__2) {$x_2$};
    \node[block] at (12,4) 
    (4_3_2__3) {$x_3$};
    \node at (12,0) {$\mathtt{g\_4.3.2.1}$};
    \draw[<-] (4_3_2__0) edge (4_3_2__1)
              (4_3_2__1) edge (4_3_2__2)
              (4_3_2__2) edge (4_3_2__3);


    \node[block] at (1,6) 
    (3_1_1__0) {$x_0$};
    \node[block] at (0,7) 
    (3_1_1__1) {$x_1$};
    \node[block] at (2,7) 
    (3_1_1__2) {$x_2$};
    \node at (1,5) {$\mathtt{g\_3.(1,1)}$};
    \draw[<-] (3_1_1__0) edge (3_1_1__1)
              (3_1_1__0) edge (3_1_1__2);

    \node[block] at (5,6) 
    (3_2__0) {$x_0$};
    \node[block] at (5,7) 
    (3_2__1) {$x_1$};
    \node[block] at (5,8) 
    (3_2__2) {$x_2$};
    \node at (5,5) {$\mathtt{g\_3.2.1}$};
    \draw[<-] (3_2__0) edge (3_2__1)
              (3_2__1) edge (3_2__2);


    \node[block] at (1,10) 
    (2__0) {$x_0$};
    \node[block] at (1,11) 
    (2__1) {$x_1$};
    \node at (1,9) {$\mathtt{g\_2.1}$};
    \draw[<-] (2__0) edge (2__1);


    \node[block] at (1,13) 
    (1__0) {$x_0$};
    \node at (1,12) {$\mathtt{g\_1}$};
    
\end{tikzpicture}
\end{center}
\caption{This figure depicts all of the isomorphism classes for Nakamoto graphs composed of 1, 2, 3, and 4 blocks.
Beneath each block state is a label for the blocks state, which includes a nested encoding helpful for extending the set to $5+$ blocks.}
\label{fig:block_states}
\end{figure}

\begin{figure}
    \centering
    \includegraphics[width=0.5\linewidth]{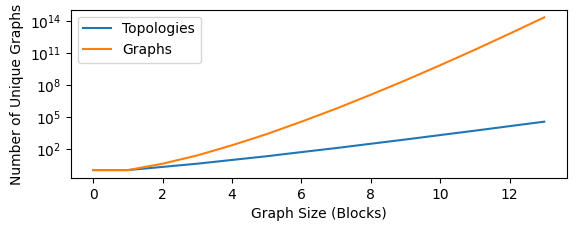}
    \caption{The number of unique graphs versus the number of unique graph topologies with respect to the number of blocks in the graph. Note that the vertical axis is log scaled.}
    \label{fig:isomorphism}
\end{figure}

As an example, Figure \ref{fig:block_states} shows the set of all isomorphism classes for Nakamoto graphs with up to 4 vertices. 
In general, the number of isomorphism classes for Nakamoto graphs with \( k \) blocks can be computed using dynamic programming. 
Figure \ref{fig:isomorphism} illustrates the benefit of modeling only isomorphism classes, which effectively removes a factorial term from the number of unique graphs (because in the worst case, blocks can be shuffled arbitrarily within each topology). 

\subsubsection{Maintaining a Finite Block Graph}
As $\mathcal{M}$ is an infinite-horizon game, there necessarily must be a mechanism by which blocks can be removed from $G_t$ to guarantee a finite state space.
We define two mechanisms to limit the number of blocks in a block graph to a maximum of $M$. 
Informally, 
\begin{enumerate}
    \item when an overwhelming majority of agents intends to append to a subgraph of $G_t$, the graph is \textbf{pruned} such that the subgraph becomes the new global block graph; and
    \item if the prune condition is not met and the number of blocks in $G_t$ would exceed $M$, the state \textbf{resets}, meaning it transitions to the initialization state with probability 1.
\end{enumerate}

More precisely, for a block graph $G_t$ and for a block $x \in G_t$, let $\Gamma(G_t,x)$ be the induced subgraph of $G_t$ containing only $x$ and its descendants.
If there exists any block $x$ such that the cumulative probability that a block in $\Gamma(G_t,x)$ is appended to is at least \( 1 - \epsilon \), where $0 \leq \epsilon < 0.5$, then $G_t$ is pruned in place so that the new block graph is $\hat{G}_t = \Gamma(G_t,x)$, and the state then transitions from $\hat{G}_t$. 
If there are multiple subgraphs which satisfy this criterion, then the smallest such subgraph is selected. 
The \emph{prune condition} for a block $x$ in a state $S_t$ is therefore defined as
\begin{equation}
    \Psi_{\bar{\pi}}(S_t,x,a_t,\mu) = 
    \begin{cases}
        1 & \frac{1}{N} \big( \mathbbm{1}_{\Gamma(G_t,x)}(a_t) + (N-1) \sum_{l} \mu(l,G_t) \mathbbm{1}_{\Gamma(G_t,x)} (\bar{\pi}(l)) \big) > 1 - \epsilon\\
        0 & \text{else},
    \end{cases}
\end{equation}
where $a_t$ is the action of the representative agent, $\epsilon$ is the \textit{prune threshold}, and $\mathbbm{1}_{\Gamma(G_t,x)}(a_t)$ is the indicator function which evaluates to $1$ if the block specified by the action $a_t$ is in $\Gamma(G_t,x)$.
When evaluating state transitions, the prune condition is checked \emph{before} any transition probabilities are computed, and then for the set of blocks
\[
\mathcal{X} = \{x: \Psi_{\bar{\pi}}(S_t,x,a_t,\mu)=1\},
\] $G_t$ is replaced with $\hat{G}_t = \Gamma(G_t,\hat{x})$, where
\[
 \hat{x} = \min_{x \in \mathcal{X}} | \Gamma(G_t,x)|.
\]
Note that because $\epsilon < 0.5$, $\hat{x}$ is a unique minimum of $| \Gamma(G_t,x)|$ because $\{\Gamma(G_t,x) : x \in \mathcal{X}\}$ is a set of nested subgraphs of $G_t$, where each subgraph is a set of blocks that the agents ``agree'' to append to under $\bar{\pi}$.

Ideally, pruning would occur frequently, meaning agents are often able to agree on a block or subgraph. 
However, this is not the case for many local policies and system states.
Thus, there must be an additional mechanism by which the block graph is kept finite: if the global block graph contains $M$ blocks and the prune condition is not met, the state \textbf{resets} to the initialization state with probability 1. 
A reset is the result of a failure to reach agreement on a block or subgraph, so from the perspective of a system designer, protocols or block selection strategies which result in frequent resetting should be avoided.

\subsection{Transition Probabilities}
\label{sec:mdp_tp}
Recall that $T_{\bar{\pi}}$ can be decomposed into component probabilities.
Below we describe how each probability is computed.

\subsubsection{Global Block Graph Transition Probability}
The probability of transitioning from \( G_t \) to \( G_{t+1} \) is defined as follows. 
Let \( \Phi(G_t,G_{t+1}) \) be the (possibly empty) set of blocks in \( G_t \) such that appending a block to any \( x \in \Phi(G_t,G_{t+1}) \) results in the block graph \( G_{t+1} \). 
For example, referring to Figure~\ref{fig:block_states},
\[
\Phi(\mathtt{g\_3.(1,1)}, \mathtt{g\_4.(2.1,1)}) = \{x_1, x_2\},
\]
because appending to either $x_1$ or $x_2$ in $\mathtt{g\_3.(1,1)}$ results in the graph topology $\mathtt{g\_4.(2.1,1)}$. 
On the other hand,
\[
\Phi(\mathtt{g\_3.(1,1)}, \mathtt{g\_4.3.2.1}) = \varnothing,
\]
because there is no single block which can be appended to in $\mathtt{g\_3.(1,1)}$ that will result in a graph of $\mathtt{g\_4.3.2.1}$. 

Then, as long as the block graph is not pruned and the number of blocks in \( G_{t+1} \) does not exceed $M$, the transition probability from \( G_t \) to \( G_{t+1} \) is given by
\begin{equation}
\label{eq:next_block_state}
    P(G_{t+1} \mid S, a_t, \mu, \bar{\pi}) = \frac{1}{N} \big( \mathbbm{1}_{\Phi(G_t, G_{t+1})}(a) + (N-1) \sum_{l} \mu(l,G_t) \mathbbm{1}_{\Phi(G_t, G_{t+1})} (\bar{\pi}(l)) \big).
\end{equation}

Computationally, $\Phi$ can be computed efficiently with a combination of a dynamic programming algorithm which takes advantage of the nested structure of a Nakamoto graph and fairly straightforward graph logic.

\subsubsection{Owned Block State Transition}
Next, we consider the probability $P(O_{t+1} \mid S_t, a_t, \mu, \bar{\pi}, G_{t+1})$.
At each state transition, exactly one new block \( x' \) is appended to the global block graph (unless the state resets, in which case the transition probabilities are already defined). 
Then there are two possible values that $O_{t+1}$ could take. 
If the representative agent generates $x'$, then $O_{t+1} = O_{t+1}^{(1)}$, which is characterized by the conditions
\[
\forall x \in G_t: O_{t+1}^{(1)}[x] = O_t[x]
\]
and
\[
O_{t+1}^{(1)}[x'] = 1.
\]
If the agent does not generate $x'$, then $O_{t+1} = O_{t+1}^{(0)}$, where
\[
\forall x \in G_t: O_{t+1}^{(0)}[x] = O_t[x]
\]
and
\[
O_{t+1}^{(0)}[x'] = 0.
\]
The probability that the representative agent generates \( x' \) is given by

\begin{equation}
P(O_{t+1}[x'] = 1 \mid s, a_t, \mu, \bar{\pi}) = 
    \frac{\mathbbm{1}_{\Phi(G,G_{t+1})}(a_t)}{\big(\mathbbm{1}_{\Phi(G,G_{t+1})}(a_t) + (N-1) \sum_{l} \mu(l) \mathbbm{1}_{\Phi(G,G_{t+1})} (\bar{\pi}(l)) \big)}.
\end{equation}

This has the following interpretation: if the RA and $m - 1$ other agents attempt to append to a block in $\Phi(G,G_{t+1})$, agent $j$ generates $x'$ with probability $\frac{1}{m}$.
If the RA does not attempt to append to a block in $\Phi(G,G_{t+1})$, then they cannot generate $x'$.

Putting things together, 
\begin{equation}
    P(O_{t+1} \mid S_t, a_t, \mu, \bar{\pi}, G_{t+1}) = 
    \begin{cases}
        P(O_{t+1}[x'] = 1 \mid S_t, a_t, \mu, \bar{\pi}) & \text{if } O_{t+1} = O_{t+1}^{(1)}\\
        1 - P(O_{t+1}[x'] = 1 \mid S_t, a_t, \mu, \bar{\pi}) & \text{if } O_{t+1} = O_{t+1}^{(0)}.
    \end{cases}
\end{equation}

\subsubsection{Received Block Vector Transition Probability}
\label{sec:transition-recv}
The blocks in $G_t$ are propagated according to independent processes, so $P(B_{t+1} \mid S_t, a_t, \mu, \bar{\pi}, G_{t+1}, O_{t+1})$ can be computed block by block. 
Let $X_{G_{t+1}}$ be the set of blocks in $G_{t+1}$. 
For any block $x$ that the agent generated or has already received as of state $S_t$, $B_{t+1}[x] = 1$ with probability 1.
For any other block $x' \in X_{G_t}$ such that $B_t[x'] = 0$ and $O_t[x'] = 0$, 
\[
P(B_{t+1}[x'] = 1 \mid S_t, a_t, \mu, \bar{\pi}, G_{t+1}, O_{t+1}) = P(H_{\alpha, \delta} = 1).
\]

Then if $z_1 = |\{x \mid x : B_t[x] = 0, B_{t+1}[x] = 1\}|$ is the total number of blocks the agent receives when the state transitions from $S_t$ to $S_{t+1}$ and $z_0 = |\{x \mid x : B_t[x] = 0, B_{t+1}[x] = 0\}|$ is the number of blocks they do not receive,
\begin{equation}
    P(B_{t+1} \mid S_t, a_t, \mu, \bar{\pi}, G_{t+1}, O_{t+1}) = P(H_{\alpha, \delta} = 1)^{z_1} (1 - P(H_{\alpha, \delta} = 1))^{z_0}.
\end{equation}

\subsection{Rewards}
Recall that the reward function takes the form 

\[
R_{\bar{\pi}}(S_t, a_t, S_{t+1}, \mu) = r Q(S_t, a_t, S_{t+1}, \mu, \bar{\pi}), 
\]
where $Q(S_t, a_t, S_{t+1}, \mu, \bar{\pi})$ is the number of critical path blocks for which the agent should receive rewards.
In most cryptocurrencies, agents are only able to spend rewards for a block after it is determined with high probability whether the block is in the critical path or not. 
This coincides neatly with our prune condition, which removes blocks from the block graph once it is sufficiently unlikely that they will be appended to again. 
Let $X(G_t,x) \subset G_t$ be the set of blocks in $G_t$ which are ancestors of $x$.
In other words, $X(G_t,x)$ is the set of blocks on the critical path that are removed from $G_t$ if $G_t$ is pruned at $x$.
Then if the block graph $G_t$ is pruned to $\hat{G}_t = \Gamma(G,x')$,

\begin{equation}
    Q(S_t, a_t, S_{t+1}, \mu, \bar{\pi}) = | \{x \in X(G_t,x') : O_t[x] = 1\}|.
\end{equation}

If the state resets because the block limit \(M\) is reached without pruning, then no rewards are granted, as $G_t$ is not pruned.
This is intentional, as it encourages agents to choose a policy which avoids resetting as much as possible. 

\subsection{Computing the Mean Field Distribution}
While it is possible to solve $\mathcal{M}_{\bar{\pi}}$ for any choice of $\mu$, in order for the solution to be a mean field equilibrium, $\mu$ must be consistent with the block propagation dynamics and symmetric across all agents (including the representative agent). 
Let the distribution $\mu_{\bar{\pi}}$ refer to the long-run average distribution over local block graphs for each agent when the non-representative agents have local policy $\bar{\pi}$. 
Solving for $\mu_{\bar{\pi}}$ analytically is challenging due to the complexity of the system dynamics and the size of the state space. 
However, because all agents in the system receive blocks according to the same processes, we can compute $\mu_{\bar{\pi}}$ as a function of the stationary distribution of $\mathcal{M}_{\bar{\pi}}$.
At steady-state, the distribution over local block graphs for the representative agent is the same as that of the non-representative agents. 

Concretely, let $f(l, G_t,\bar{\pi},\mu)$ be the steady-state probability of local block graph $l$ for the representative agent when the global block graph is $G_t$.
Then a mean field distribution $\mu_{\bar{\pi}}$ is stationary if and only if $\mu_{\bar{\pi}}$ is a fixed point of $f$, meaning for every $l,G_t \in \mathcal{G}$,
\[
\mu_{\bar{\pi}}(l,G_t) = f(l,G_t,\bar{\pi},\mu_{\bar{\pi}}).
\]
Computing $\mu_{\bar{\pi}}$ therefore reduces to computing the fixed point of $f$, which can be accomplished with the following iterative algorithm. 

First, determine an initial estimate $\mu^0_{\bar{\pi}}$. 
There are a number of ways to estimate $\mu^0_{\bar{\pi}}$, but in our experiments we use the following estimation. 
For any global block graph $G_t$ and some (non-representative) agent $i$, let 
\[
p_D(G_t,x) = P\bigl(H_{\alpha,\delta} \leq D(G_t, x)+1\bigr),
\]
and 
\[
p_C(G_t,x) = P\bigl(H_{\alpha,\delta} \leq C(G_t, x)+D(G_t,x)+1\bigr),
\]
where $D(G_t,x)$ is the number of descendants of $x$ in $G_t$ and $C(G_t,x)$ is the number of blocks in $G_t$ which are neither descendants nor ancestors of $x$. 
Then the probability that some block $x \in G_t$ is in agent $i$'s local block graph is bounded by 
\[
p_D(G_t,x) \leq P(x \in L^i \mid G_t) \leq  p_C(G_t,x).
\]
This is because $P(x \in L^i \mid G_t)$ depends on the number of block steps which have elapsed since $x$ was generated, which can be bounded using fairly straightforward reasoning on the order in which $G_t$ was constructed.
$D(G_t,x)$ is the number of blocks that must have been generated \textit{after} $x$, as they are descendants of $x$.
On the other hand, $C(G_t,x)$ is the number of blocks in $G_t$ for which the generation order with respect to $x$ cannot be determined, as they might have been generated before or after $x$. 
Then for each possible local block graph $l$ and global block graph $G_t$, we set
\[
\mu^0_{\bar{\pi}}(l,G_t) = \prod_{x \in l} \frac{1}{2} (p_D(G_t,x) + p_C(G_t,x)),
\]
i.e., the probability of each local block graph is initialized to be halfway between the lower and upper bounds. 

After an initial estimate $\mu^0_{\bar{\pi}}$ has been determined, the next step is to solve  $\mathcal{M}_{\mu^0_{\bar{\pi}}}$ and determine the resulting steady-state distribution. 
For all $l, G_t \in \mathcal{G}$,
\[
\mu^1_{\bar{\pi}}(l,G_t) = f(l,G_t,\bar{\pi},\mu^0_{\bar{\pi}}).
\]
Subsequently, $f(l,G_t,\bar{\pi},\mu^1_{\bar{\pi}})$ is computed and this procedure is iterated until a fixed point of $f$ is achieved, at which point $\mu_{\bar{\pi}}$ has been determined.

\subsection{Computing the Mean Field Equilibrium}
\label{sec:compute_mfe}
Recall from Section \ref{sec:eq_c} that the mean field equilibrium of $\mathcal{M}$ is characterized by a fixed point of the Bellman equation 
\[
\bar{\pi}_j^*(L^j_t) = \arg \max_a \sum_{s} T_{\bar{\pi}^*}(S_t,a,s,\mu_{\bar{\pi}})[R_{\bar{\pi}^*}(S_t,a,s,\mu_{\bar{\pi}}) + \gamma V^*(s)]
\]
and
\[
V^*(s) = \max_a \sum_{s'} T_{\bar{\pi}^*}(s,a,s',\mu_{\bar{\pi}})[R_{\bar{\pi}^*}(s,a,s',\mu_{\bar{\pi}}) + \gamma V^*(s')].
\]
However, such a $\bar{\pi}_j^*(L^j_t)$ is not guaranteed to exist, as there may be multiple states which correspond to the same local block graph for the RA, and the optimal action may not be the same for each state.
If the set of all states which evoke the same local block graph as $S_t$ is $\sigma(S_t)$, then an additional constraint must be placed upon $\bar{\pi}_j^*$ such that 
\[
\forall s, s' \in \sigma(S_t) : \bar{\pi}_j^*(s) = \bar{\pi}_j^*(s').
\]
We describe two logical approaches to computing $\bar{\pi}_j^*$ under this constraint. 

\textit{Approach 1}. The first is that the definition of $\bar{\pi}_j^*(L^j_t)$ could be modified to accommodate the additional constraint. 
Rather than selecting each action to maximize the expected future reward of the current state, the representative agent could instead maximize some function (such as the L1 norm) of the expected future reward over all possible states which evoke the local block graph $L^j_t$.
This means solving
\[
\bar{\pi}_j^*(L^j_t) = \arg \max_a \big\| \sum_{s} T_{\bar{\pi}^*}(\hat{s},a,s,\mu_{\bar{\pi}})[R_{\bar{\pi}^*}(\hat{s},a,s,\mu_{\bar{\pi}}) + \gamma V^*(s)]\big\|_{\hat{s} \in \sigma(S_t)}
\]
and
\[
V^*(s) = \max_a  \big\| \sum_{s'} T_{\bar{\pi}^*}(\hat{s},a,s',\mu_{\bar{\pi}})[R_{\bar{\pi}^*}(\hat{s},a,s',\mu_{\bar{\pi}}) + \gamma V^*(s')]\big\|_{\hat{s} \in \sigma(s)}.
\]
The representative agent is then guaranteed to select the same action for each state $\hat{s} \in \sigma(S_t)$. 
The intuition behind this approach is that the agent recognizes that there are a number of system states which might lead to the local block graph they observe, so rather than choosing the best action for any specific state, they optimize against some distribution over each possible state (in the case of the L1 norm, this is the uniform distribution).

\textit{Approach 2.} The second approach involves solving $\mathcal{M}_{\bar{\pi}^*}$ as though it were a fully observable MDP and converting the resulting (full) policy into a local policy. 
This simply means solving 
\[
\pi_j^*(S_t) = \arg \max_a \sum_{s} T_{\bar{\pi}^*}(S_t,a,s,\mu_{\bar{\pi}^*})[R_{\bar{\pi}^*}(S_t,a,s,\mu_{\bar{\pi}^*}) + \gamma V^*(s)]
\]
and
\[
V^*(s) = \max_a \sum_{s'} T_{\bar{\pi}^*}(s,a,s',\mu_{\bar{\pi}^*})[R_{\bar{\pi}^*}(s,a,s',\mu_{\bar{\pi}^*}) + \gamma V^*(s')]
\]
and then deriving $\bar{\pi}_j^*$ from $\pi_j^*$ using the following method. 
At time $t$, when the RA has local block graph $L^j_t$, let $S'$ be a state such that $G' = L^j_t$ and $\forall x \in G': R'[x] =1$.
Then a local policy can be derived from a full policy by 
\[
    \label{eq:full_to_local}
    \bar{\pi}_j(L^j_t) = \pi_j(S').
\]
Put into words, the action agent $j$ takes under the local policy for local block graph $L^j_t$ is the action they would take under the full policy if the state were $S'$, where the global block graph of $S'$ is $L^j_t$ and the agent has received every block. 
Intuitively, this reflects the practical notion that a cryptocurrency miner chooses actions under the assumption that the local block graph they observe \textit{is} the global block graph.

In practice, the strategy that most cryptocurrency miners follow aligns far more closely with Approach 2, as mining software customarily does not consider unreceived blocks at all when deciding which block to append to. 
As such, in our experiments we implement Approach 2, and we defer exploration of Approach 1 to future research.

\section{Implementation Details}
\label{sec:implementation}
We implement our framework in Python3 in order to examine the strategic behavior of cryptocurrency miners and analyze the structure and evolution of the resulting block graph.
In order to provide intuition on the practical implications of each system parameter, we begin with an exploratory analysis of $\mathcal{M}$. 
Subsequently, we describe how policies are defined in our implementation. 

\subsection{System Parameters}

Our framework takes as input a number of parameters corresponding to varying system dynamics. 
These parameters are summarized in Table \ref{table}, and unless otherwise specified, we use the default value in our experiments. 
In order to provide insights that might be helpful for understanding or extending our work, we briefly discuss the impact of each parameter on miner behavior and system dynamics.

\begin{table}
    \centering
    \begin{tabular}[c]{ |p{5em}|p{22em}|p{8em}| } 
 \hline
 \textbf{Parameter} & \textbf{Description} & \textbf{Default Value} \\ 
 \hline
 \centering $N$ & The number of agents. & $1000$ \\ 
 \hline
 \centering $M$ & The maximum number of blocks in the global block graph. & $5$ \\ 
 \hline
 \centering $\alpha$ & The probability that a block is generated in a single time step. & $0.001$ \\ 
 \hline
 
\centering $\delta$ & For any block that an agent has not yet received, the probability that they receive it in the next time step is $\delta$. & $0.01$  \\ 
 \hline

\centering $\gamma$ & The discount rate on future rewards. & $0.99$  \\ 
 \hline

\centering $\epsilon$ & The prune threshold is $1 - \epsilon$, meaning if the probability that a subgraph of $G_t$ will be appended to is greater than $1 - \epsilon$, then the block graph is immediately pruned. & $0.01$  \\ 
\hline

\centering $r$ & The reward that the representative agent receives for generating a critical path block. & $1$\\ 
\hline

\centering $\bar{\pi}_0$ & The initial policy for the non-representative agents. & LCR  \\ 
\hline

\end{tabular}
    \caption{Summary of system parameters.}
    \label{table}
\end{table}

\subsubsection{Number of Agents} 
Under the mean field approximation, the number of agents is assumed to be large enough that the impact each individual agent has on the state is small. 
In general, the only way that the number of agents impacts the equilibrium policy is through the probability that the representative agent generates each block (which is $1/N$, as agents are assumed to have identical capabilities). 
Therefore $N$ impacts only the magnitude of the rewards that the agent receives, but not their choice of action, so all of our experiments use the default value of $N=1000$. 

\subsubsection{Maximum Block Graph Size} 
Using our implementation, we are able to solve an MDP with a global block graph of up to $M=7$ blocks within a reasonable amount of (wall clock) time, which corresponds to $54053$ unique states. 
However, we observe nearly identical behavior for block graphs of $4, 5, 6$, and $7$ blocks, so most of our experiments are performed with a maximum block graph size of $5$, which corresponds to 1537 states.
In practice, it is rare for a miner to have to choose between more than 5 blocks, making $M=5$ a reasonable default parameter. 

\subsubsection{Block Generation and Propagation}
Individually, the parameters $\alpha$ and $\delta$ could be set arbitrarily, as the duration of a time step is not fixed. 
Selecting $\alpha = 0.001$ could correspond to a rate of one block generated per nanosecond or one block per century, depending on the duration of a time step. 
However, the choice of $(\alpha, \delta)$ as a pair is far more significant, as it corresponds to the relative rate at which blocks are generated and propagated. 
In our experiments, a ratio of $\frac{\delta}{\alpha} = 10$ yields stable results, although it is not the only ratio to do so. 
This might might correspond to, for example, an expected block time of 10 minutes and an expected propagation delay of 1 minute. 

\subsubsection{Discount Factor}
The choice of $\gamma$ does not significantly impact the equilibria we observe. 
In practice, the discount factor should be assumed to be fairly close to 1, as the time between blocks is generally small enough that miners place the same value on receiving a reward for one block as they would for the next block. 

\subsubsection{Prune Threshold}
In general, the choice of $\epsilon$ corresponds to the level of certainty required before a block is determined to be in the critical path or not.
Practically, once a block is pruned from the global block graph, it can be considered finalized, as it can no longer be appended to or removed from the block graph. 
The choice of prune threshold might therefore be of independent interest in the study of settlement and finality in Nakamoto-style cryptocurrencies. 
However, we defer analysis of $\epsilon$ to future work and utilize a default value of $\epsilon = 0.01$ in our experiments. 
This has the interpretation that any block which will be appended to with a probability of less than $0.01$ is pruned from the block graph. 

\subsubsection{Rewards}
The magnitude of the reward the representative agent receives for generating a block is not individually meaningful as long as it is positive, so for convenience we set $r = 1$ in all experiments. 

\subsubsection{Initial Policy}
The equilibria we observe depend heavily on the initial policy followed by the NRAs.
This is because the representative agent only receives a reward if the block graph is pruned. 
For initial policies which do not result in pruning, there is no policy the representative agent can follow which will yield a positive reward. 
For initial policies which do result in pruning, the representative agent attempts to choose actions which maximize i) the probability that pruning occurs and ii) the probability that they generate critical path blocks. 
This aligns neatly with practical intuition--if all miners simply choose a block to append to at random, it is unlikely that a meaningful critical path will form and it therefore does not matter what the RA does. 
For a policy that does yield a meaningful critical path (such as the LCR), it is in the RA's best interests to follow the same policy.

\subsection{Implementing a Policy}
A full policy is a mapping from a state to an action, and a local policy is a mapping from a local block graph to an action. 
Each action corresponds to a specific block in the agent's local block graph.
While this definition is sufficient in theory, practical issues arise when implementing such a policy. 
Below we discuss these issues.  

\subsubsection{Agents With an Empty Local Block Graph}
Recall that the local block graph of an agent who has not received the root block of $G_t$ is the null graph. 
Such an agent has no available actions, and as a result they cannot append a block to $G_t$. 
This impacts the probability of transitioning from $G_t$ to $G_{t+1}$, as only a subset of the $N$ agents are able to generate blocks at time $t$. 
To address this issue, we normalize Equation \ref{eq:next_block_state} with a multiplicative factor $\kappa$, where $\frac{1}{\kappa}$ is the fraction of agents with no blocks in their local block graph (potentially including the RA). 

\subsubsection{Topologically Identical Blocks}
Another issue arises when implementing a deterministic strategy--there must be a deterministic way to choose between \textit{topologically identical} blocks which can have their position in the graph swapped without changing the topology of the graph. 
For example, consider the graph $\mathtt{g\_4.(1,1,1)}$ from Figure \ref{fig:block_states}. 
The blocks $x_1, x_2$, and $x_3$ are all topologically identical, as their positions can be swapped without changing the topology of the graph.
In order to determine a fixed point of the MDP, there must be a way by which any agent with the same local block graph chooses the same block. 

To address this issue, we assume that although blocks are functionally identical, they can still be differentiated from one another. 
Agents then all follow a deterministic rule to select between topologically identical blocks, such as always selecting the block with the lowest block hash or, in the case of our visual example, always selecting the ``leftmost'' block.

\section{Results}
\label{sec:perf}
In order to demonstrate the utility of our framework, we use it to prove two main results. 
First, we study the impact that varying the block propagation delay has on PoW efficiency--the fraction of blocks which end up in the critical path.
Our observations align closely with theoretical results describing the relationship between network delay and PoW Efficiency.
This alignment validates our usage of the mean field approximation, as the stationary behavior of our framework is consistent with expected theoretical results.

Next we investigate the mean field equilibria of $\mathcal{M}$.
Using a proof by exhaustive search, we find that the LCR is the optimal block selection strategy for a Nakamoto-style cryptocurrency, as it has the highest stationary PoW efficiency of any equilibrium policy.
However, we found the longest chain rule to be quite fragile, as the equilibrium determined by our implementation only converges upon the LCR when the initial policy of the non-representative agents is quite similar to the LCR.

\subsection{Network Delay and PoW Efficiency}
In asynchronous blockchain protocols, the time it takes for a block to be propagated to the network (the network delay) can significantly impact the the \textit{PoW efficiency} of the protocol--the fraction of blocks which end up in the critical path. 
This is because as the network delay increases, the amount of time that agents spend mining on inconsistent local block graphs also increases, resulting in a greater likelihood of non-critical path blocks. 
Theoretical results show that the PoW efficiency of a blockchain following the LCR is 
\[
\frac{1}{1 + \alpha \Delta},
\] 
where $\Delta$ is the end-to-end network delay \cite{georghiades2022scalable, pass2017analysis}. 
This theoretical relationship can be used to test how well the mean field approximation we adopt is able to replicate practical system behavior. 
To this end, we aim to prove the following claim. 

\begin{claim}
\label{claim:eff}
    Under the mean field approximation, the PoW efficiency of a blockchain following the LCR is given by 
    \[
\frac{1}{1 + \alpha \Delta},
\] 
where $\alpha$ is the block generation rate and $\Delta$ is the network delay.
\end{claim}

To test this claim, we adopt the default values described in Table \ref{table} for each parameter except for $\delta$, which we vary in order to study the impact of network delay on PoW efficiency.
For each value of $\delta$, the network delay $\Delta$ can be computed as the number of time steps required for a $\rho$ fraction of the agents to receive a block. 
Note that in our model, the number of time steps required to guarantee that all agents have received a block is infinite, so we model the PoW efficiency for a few different values of $\rho < 1$ which correspond finite network delays. 

\begin{figure}
    \centering
    \includegraphics[width=0.8\linewidth]{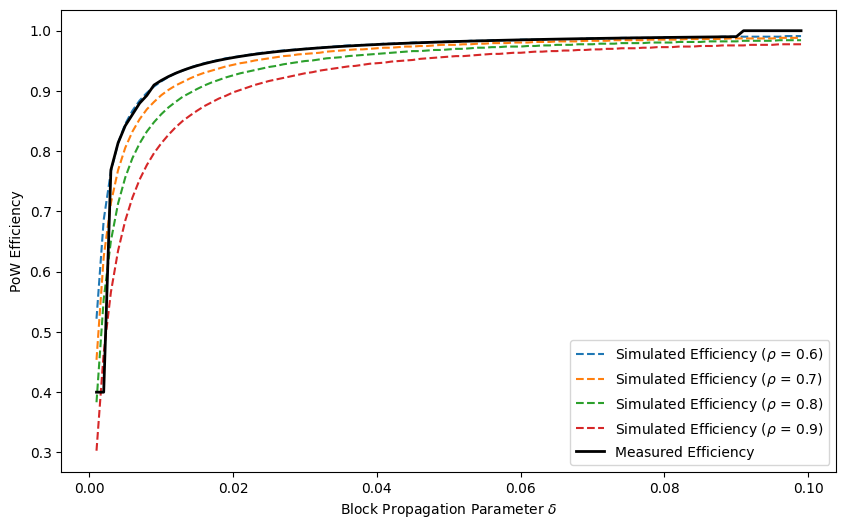}
    \caption{The rate at which non-critical path blocks are appended to the block graph with respect to the block propagation parameter $\delta$. The dashed lines correspond to the theoretical mining efficiency for different choices of network delay, while the solid line is the mining efficiency we measure at equilibrium.}
    \label{fig:propagation_delay}
\end{figure}

For each value of $\rho$, we plot the theoretical PoW efficiency against the stationary PoW efficiency of $\mathcal{M}$ under the LCR equilibrium policy. 
As shown in Figure \ref{fig:propagation_delay}, the PoW efficiency of $\mathcal{M}$ at equilibrium aligns closely with the expected result, constituting a proof of Claim \ref{claim:eff}.

\subsection{Equilibrium Optimality}
We find that the LCR is not a unique mean field equilibrium of $\mathcal{M}$.
In fact, the equilibria determined by our framework are highly susceptible to changes in the initial policy $\bar{\pi}_0$.
In general, initial policies which do not result in frequent pruning yield equilibria with the same characteristic.
This is because in the absence of pruning, there is no policy the RA could choose that would yield a non-zero reward, and they are therefore indifferent between actions for almost any state.
In this case, all possible strategies are weakly dominant, as the RA receives a reward of zero regardless of their actions.

Given that there are multiple mean-field equilibria of $\mathcal{M}$, the question of equilibrium quality arises.
There are many possible metrics by which to compare equilibria, but we focus specifically on PoW efficiency, which has direct ties to security, consensus stability, and energy efficiency in PoW blockchains \cite{georghiades2022scalable, pass2017analysis}.
Intuition suggests that the LCR should be the block selection strategy which optimizes PoW efficiency in Nakamoto-style blockchains because i) rewards are typically only granted for blocks that end up in the critical path, and ii) a reasonable heuristic for generating a block in the critical path is to append to the current critical path. 
Some degree of implicit empirical evidence also supports this conclusion, as Bitcoin and other Nakamoto-style cryptocurrencies have been operating under the LCR for many years. 
However, to the best of our knowledge, optimality of the LCR as a block selection strategy has not been proven. 
Our framework supports such a proof. 
Concretely, we aim to prove the following claim. 

\begin{claim}
\label{claim:opt}
    Under the assumptions in Section \ref{sec:mdp}, the LCR is the mean field equilibrium policy which maximizes the stationary PoW efficiency of the induced Markov chain. 
\end{claim}

We prove this claim by exhaustive search. 
That is, we compute every possible mean field equilibrium and compute the resulting PoW efficiency at steady state. 
For this, we limit the size of the block graph to 4 blocks\footnote{The number of unique policies grows super-exponentially with the size of the block graph. For a maximum block graph of size 5, there are 165,888,000 unique policies, making an exhaustive search impractical for any block size greater than 4.}, which corresponds to 1152 unique policies.
Of these, 150 policies are mean field equilibria of $\mathcal{M}$, although the majority of these equilibria are weakly dominated and result in zero reward for the RA. 

\begin{figure}
    \centering
    \includegraphics[width=0.75\linewidth]{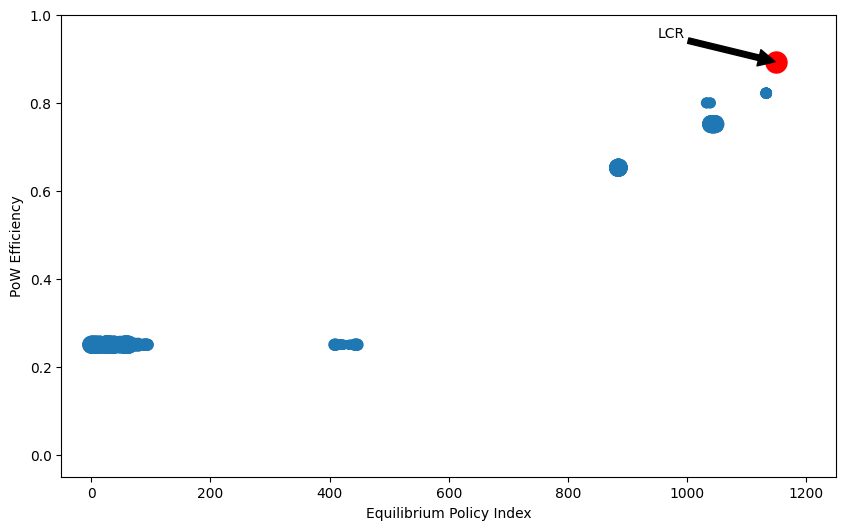}
    \caption{The PoW efficiency of each equilibrium policy of $\mathcal{M}$. The horizontal axis indicates the index of the equilibrium policy and can be ignored. The size of each marker corresponds to the frequency with which the equilibrium policy is encountered.}
    \label{fig:prune_scatter}
\end{figure}

Figure \ref{fig:prune_scatter} shows the PoW efficiency for each possible equilibrium policy, with the size of the markers indicating the frequency with which the equilibrium is encountered.
We find that the LCR is uniquely optimal as an equilibrium policy, and that policies that are more similar to the LCR achieve higher PoW efficiency than policies which vary significantly from the LCR. 
As there are no possible equilibrium policies that might surpass the PoW efficiency of the LCR, this constitutes a proof of Claim \ref{claim:opt}.

\section{Future Work and Conclusion}
\label{sec:conclusions}





We identify several interesting directions in which $\mathcal{M}$ can be extended to support different types of analysis, and we follow with a few concluding remarks. 

\subsection{Future Work}
\textbf{Alternate Reward Functions.} 
Reward functions play a fundamental role in shaping the incentives of miners. 
Our framework can aid in the study of how modifications to the reward structure, such as rewarding agents for non-critical path blocks or partial shares, affect equilibrium strategies and system dynamics. 
It has been theorized that blocks containing sufficiently high transaction fees may cause inefficiencies in the consensus process--this could be examined under our framework by adding state variables which indicate whether a block has ``high'' fees, ``low'' fees, or anywhere in between. 

\textbf{Non-uniform Distribution of Mining Power.} 
In our model, all agents have equal mining power. 
This symmetry allows us to solve for a mean field equilibrium for any set of system parameters. 
In real-world cryptocurrency systems, mining power is not uniformly distributed among all participants, and multiple types of attacks are only viable once the attacker crosses a certain threshold of the total mining power. 
Our framework can handle the scenario in which the representative agent has a larger share of the mining power by modifying only the transition probability function. 
In this case, our framework will allow for computation of the representative agent's best response to any NRA policy, but as the agents are no longer symmetric it will not be able to search for equilibrium policies. 

\textbf{Withholding Attacks.} Withholding attacks are strategies in which a miner withholds blocks from the network in order to gain a strategic advantage.
The two most prominent withholding attacks are selfish mining and the private double-spend. 
In our framework, these can be simulated by allowing the representative agent to withhold blocks until certain conditions are met. 
Static withholding attacks (which do not adapt to evolving system dynamics) could be simulated with little or no modification to the state space, as the withholding strategy and payoff could be instilled within the transition probability function and reward function. 

\textbf{Alternate Blockchain Structures.} 
While our framework is based on Nakamoto graphs, blockchains based on other graph structures, such as Directed Acyclic Graphs (DAGs) or parallel chains, are also of significant interest. 
Implementing such a structure would require a modification of the state space and transition probability function to accommodate the different graph topologies, while the system dynamics and solution method would remain the same. 

\textbf{Alternate Block Propagation Models.} 
In many blockchain systems, different types of agents may receive blocks at different rates based on network topology, geographical distance, or bandwidth restrictions.
Similarly, there are a variety of broadcast algorithms which might be used to disseminate blocks. 
Studying the impact of these network dynamics could be accomplished by adding additional mean field variables to the model and adding state variables to indicate how different types of blocks should be propagated. 

\textbf{Alternate Optimality Conditions}
In Section \ref{sec:perf}, we find that the LCR is optimal in maximizing the PoW efficiency. 
However, there are other metrics by which the quality of a block selection strategy can be quantified, such as settlement time, attack resistance \cite{zhang2019lay}, fairness in reward distribution, etc.
Indeed, most protocols which build upon Bitcoin are designed precisely to improve upon one or more of these metrics. 
In general, as our framework enables computation of a complete transition probability matrix for each equilibrium policy, these metrics could be characterized at steady state for any protocol which can be represented with sufficient modification to the block structure or transition probability function of $\mathcal{M}$.
In the case of multiple equilibrium policies, this provides a straightforward and scalable method for determining the optimal block selection strategy for any protocol and any optimality condition.

\subsection{Conclusion}
Blockchains have become a foundational technology for decentralized systems, but evaluating and improving their performance and security remains a complex challenge. 
In this work, we propose a framework for modeling blockchain dynamics that allows us to compute the mean field equilibrium of the blockchain growth game by solving a sequence of POMDPs. 
In doing so, we provide a structured and scalable approach to analyzing consensus mechanisms and miner behavior.
Our framework allows for the rigorous evaluation of different system dynamics, reward structures, and block selection strategies, offering insights into how these factors impact blockchain performance at steady state. 

We study the tradeoff between network delay and PoW efficiency and find that the stationary PoW efficiency we observe in our framework aligns closely with theoretical understanding. 
This result validates our use of the mean field approximation and solution methodology. 
Additionally, we present the first proof of optimality of the LCR in Nakamoto-style blockchains. 
Through exhaustive search, we show that the LCR induces the highest stationary PoW efficiency of any equilibrium policy, justifying its continued dominance in these systems.
While focused on Nakamoto-style blockchains, our model is adaptable to other blockchain architectures, making it a versatile tool for future research aimed at improving the efficiency, security, and performance of blockchain systems.

\bibliographystyle{ACM-Reference-Format}
\bibliography{References}










\end{document}